\documentclass[12pt]{article} 

\usepackage{bbding}
\usepackage{mathtools}
\PassOptionsToPackage{normalem}{ulem}
\usepackage{ulem}

\usepackage{tikz,tikz-3dplot}
\usepackage{xcolor}
\usetikzlibrary{positioning}
\usetikzlibrary{3d}

\usepackage{epsfig}
\usepackage{graphicx}
\usepackage{comment}
\usepackage{latexsym}
\usepackage{hyperref}
\usepackage{amsmath}
\usepackage{color}
\usepackage{amsbsy}
\usepackage{amssymb}
\usepackage{amsthm}
\usepackage{amsfonts}
\usepackage{cite}

\newcommand{\Z}{\mathbb{Z}}

\renewcommand{\O}{{\cal O}}

\newcommand{\diag}{{\rm diag}\,}

\newcommand{\ie}{{\em i.e.} }

\newcommand{\where}{\mbox{where}}
\newcommand{\with}{\mbox{with}}

\renewcommand{\and}{\mbox{and}}

\newcommand{\N}{{\cal N}}

\newcommand{\W}{{\cal W}}

\newcommand{\G}{{\cal G}}

\newcommand{\V}{{\cal V}}

\newcommand{\Hc}{{\cal H}}
\renewcommand{\SS}{Scherk-Schwarz }

\newcommand{\half}{{1\over 2}}

\newcommand{\be}{\begin{equation}}
\newcommand{\ee}{\end{equation}}
\newcommand{\dis}{\displaystyle}

\newcommand{\bm}{\boldmath} 

\catcode`\@=11
\def\marginnote#1{}
\newcount\hour
\newcount\minute
\newtoks\amorpm
\hour=\time\divide\hour by60 \minute=\time{\multiply\hour by60
\global\advance\minute by-\hour}
\edef\standardtime{{\ifnum\hour<12 \global\amorpm={am}%
        \else\global\amorpm={pm}\advance\hour by-12 \fi
        \ifnum\hour=0 \hour=12 \fi
        \number\hour:\ifnum\minute<10 0\fi\number\minute\the\amorpm}}
\edef\militarytime{\number\hour:\ifnum\minute<10 0\fi\number\minute}
\def\draftlabel#1{{\@bsphack\if@filesw {\let\thepage\relax
   \xdef\@gtempa{\write\@auxout{\string
      \newlabel{#1}{{\@currentlabel}{\thepage}}}}}\@gtempa
   \if@nobreak \ifvmode\nobreak\fi\fi\fi\@esphack}
        \gdef\@eqnlabel{#1}}
\def\@eqnlabel{}
\def\@vacuum{}
\def\draftmarginnote#1{\marginpar{\raggedright\scriptsize\tt#1}}
\def\draft{\oddsidemargin -.2truein
        \def\@oddfoot{\sl preliminary draft \hfil
        \rm\thepage\hfil\sl\today\quad\militarytime}
        \let\@evenfoot\@oddfoot \overfullrule 3pt
        \let\label=\draftlabel
        \let\marginnote=\draftmarginnote
   \def\@eqnnum{(\theequation)\rlap{\kern\marginparsep\tt\@eqnlabel}%
\global\let\@eqnlabel\@vacuum}  }
\def\thebibliography#1{
\vskip 0.5cm \centerline{\bf \Large References}
\list{
[\arabic{enumi}]}{\settowidth\labelwidth{[#1]}
\leftmargin\labelwidth
\advance\leftmargin\labelsep
\usecounter{enumi}}
\def\newblock{\hskip .11em plus .33em minus .07em}
\sloppy\clubpenalty4000\widowpenalty4000
\sfcode`\.=1000\relax}

\renewcommand{\theequation}{\arabic{section}.\arabic{equation}}
\renewcommand{\section}{\setcounter{equation}{0}\@startsection
{section}{1}{0mm}{-\baselineskip}{0.5\baselineskip} {\normalfont\Large\bfseries}}
\renewcommand{\subsection}{\@startsection
{subsection}{2}{0mm}{-\baselineskip}{0.5\baselineskip} {\normalfont\large\bfseries}}
\renewcommand{\subsubsection}{\@startsection
{subsubsection}{3}{0mm}{-\baselineskip}{0.5\baselineskip}
{\normalfont\normalsize\slshape}}

\begin{document}


\begin{titlepage}
\begin{flushright}
CPHT-PC001.012019, January  2019
\vspace{1.5cm}
\end{flushright}
\begin{centering}
{\bm\bf \Large Quantum stability in open string theory \\with  broken supersymmetry\footnote{Based on work done in collaboration with S.~Abel, E.~Dudas and  D.~Lewis~\cite{stableWL}, presented at DISCRETE18, Vienna,  26-30 November 2018.}}

\vspace{5mm}

 {\bf Herv\'e Partouche\footnote{herve.partouche@polytechnique.edu}}

 \vspace{1mm}

{Centre de Physique Th\'eorique, Ecole Polytechnique,  CNRS\footnote{Unit\'e  mixte du CNRS et de l'Ecole Polytechnique, UMR 7644.}, \\ Route de Saclay, 91128 Palaiseau, France}

\end{centering}
\vspace{0.1cm}
$~$\\
\centerline{\bf\Large Abstract}\\

\begin{quote}

\hspace{.6cm} We consider the 1-loop effective potential in type I string theory compactified on a torus, with supersymmetry broken by the \SS mechanism. At fixed supersymmetry breaking scale $M$, and up to exponentially suppressed terms, we show that the potential admits local minima of arbitrary sign, in dimension $d\le 5$. While the open string Wilson lines are massive, the closed string moduli are flat directions. In a T-dual picture, the relevant backgrounds involve isolated $\half$-branes, whose positions are frozen on orientifold planes, thus decreasing the rank of the gauge group, and introducing massless fermions in fundamental representations.   

\end{quote}




\end{titlepage}
\newpage
\setcounter{footnote}{0}
\renewcommand{\thefootnote}{\arabic{footnote}}
 \setlength{\baselineskip}{.7cm} \setlength{\parskip}{.2cm}

\setcounter{section}{0}


\section{Introduction}

In string theory, when supersymmetry is spontaneously broken in flat space at a scale $M$ moderately smaller than the string scale $M_s$, the effective potential simplifies greatly and the questions of its sign, magnitude  and stability at extrema can be addressed is a systematic way. In this work, we focus on the type IIB orientifold theory compactified on a torus $T^{10-d}$ of metric $G_{IJ}$, with supersymmetry spontaneously broken \`a la Scherk-Schwarz~\cite{open2_ss} along the internal direction $X^9$. In this case, $M=M_s\sqrt{G^{99}}\ll M_s$, where $G^{IJ}=G_{IJ}^{-1}$. Note that because of the underlying extended supersymmetry (with 16 supercharges), all moduli fields can be interpreted as Wilson lines (WLs), but we find convenient to use this denomination for all moduli fields except the so-called no-scale modulus $M$~\cite{noscale}.  

The dominant contribution of the 1-loop effective potential $\V$ arises from the lightest states of the spectrum. Assuming that the background does not contain mass scales between 0 and $M$, the latter are $n_F$ and $n_B$ massless fermionic and bosonic degrees of freedom, with their Kaluza-Klein (KK) states propagating along the large direction~$X^9$. The result takes the form 
\be
\V\simeq (n_F-n_B)\,\xi_d\, M^d,
\label{vres}
\ee
where $\xi_d>0$ captures the contributions of the KK modes~\cite{Itoyama:1986ei}. At such a point in moduli space, the potential is also critical with respect to the WLs, as follows from the enhancement of the massless spectrum (no Higgs-like scale between 0 and $M$). However, even if all tadpoles vanish (except for $M$ when $n_F\neq n_B$), the stability of the background is guaranteed only if the WLs are non-tachyonic, when $\V$ is expanded at quadratic order in small marginal deformations. The point is that the higher $\V$ is (due to the presence of massless fermions), the more unstable the background is, as follows from dangerous contributions arising from the massless fermions charged under the gauge group \ie coupled to the WLs. In the present work, we review the fact  that stable backgrounds exist, while satisfying $\V\ge 0$~\cite{stableWL}. 

Note that such models may be relevant in several respects. Nearly vanishing potentials may be relevant for achieving the goal of describing a small (and positive) cosmological term. Moreover,  flat cosmological evolutions of models with positive potentials have been shown to be attracted to a ``Quantum No-Scale Regime'' describing an expanding universe, where the no-scale structure, which is exact classically, is restored at the quantum level. On the contrary,  when the potential can reach negative values, the universe eventually collapses, unless the initial conditions are tuned in a tiny region of the phase space~\cite{CFP, CP}. 

We describe the  models in a geometric picture obtained by T-dualizing all internal directions. $\half$-branes with positions frozen on orientifold planes play a crucial role, since 
$(i)$ they do not yield marginal deformations that may be tachyonic, 
$(ii)$ they decrease the dimension of the gauge group, thus lowering $n_B$,
$(iii)$ and they yield massless fermionic strings stretched between them and other branes, due to the interplay between WL deformations and \SS mechanism.  


\section{Wilson line stability in 9 dimensions}

The main ideas  to raise minima of the effective potential can be understood in 9 dimensions. In  type I theory, the gauge group arising from open strings is actually $O(32)$ rather than $SO(32)$. Hence, two disconnected open string WLs moduli spaces can be considered, which are {\em disconnected} to one another, because parametrized by WL matrices of determinant 1 or $-1$~\cite{Schwarz:1999xj},  
\be
\begin{aligned}
\W &= \diag(e^{2i\pi a_\alpha}, \alpha=1,\dots, 32) \\
& \equiv \begin{cases}\;\;\;\,\: \diag(e^{2i\pi a_1},e^{-2i\pi a_1}, \dots, e^{2i\pi a_{16}},e^{-2i\pi a_{16}})  \\ \mbox{or }\diag(e^{2i\pi a_1},e^{-2i\pi a_1}, \dots, e^{2i\pi a_{15}},e^{-2i\pi a_{15}},1,-1).
\end{cases}
\end{aligned}
\label{wlma}
\ee
In both cases, the open strings having Chan-Paton charges at their ends, their KK momentum along $S^1(R_9)$ takes  the form 
\be
 {m_9+{F\over 2}+a_\alpha-a_\beta\over R_9}\,M_s,
 \label{wls}
\ee
where $m_9\in\Z$, $F$ is the fermionic number and $R_9=\sqrt{G_{99}}\gg 1$, where in our conventions all moduli fields are dimensionless. It is convenient to switch to the type I' picture obtained by T-dualizing $R_9\to 1/R_9=\tilde R_9$. The internal space becomes $S^1(\tilde R_9)/\Z_2$, with two orientifold O8-planes located at the fixed points $\tilde X^9=0$ and $\tilde X^9=\pi \tilde R_9$. In this framework, the WLs are the positions $\tilde X^9=2\pi a_\alpha$ of 32 $\half$-D8-branes. When two $\half$-branes are located at $2\pi a_\alpha$ and $-2\pi a_\alpha$ on the double cover, they are actually coincident in $S^1(\tilde R_9)/\Z_2$ and give rise to a plain D8-brane, whose position $a_\alpha$ is a modulus free to vary. On the contrary, when a single $\half$-brane is not paired with a mirror object, a fact that can only happen when $a_\alpha=0$ or $\half$, it is frozen on one of the O8-planes and we are left with only 15 independent WL deformations, as indicated in the last line of Eq.~(\ref{wlma}). When there are $p_1$ $\half$-branes located at $a=0$, $p_2$ at $a=\half$, and stacks of $r_\sigma$ $\half$-branes at $a_\sigma\!\in\,) 0,\half($ with their mirrors at $-a_\sigma$, the gauge symmetry is $U(1)_{G}\times U(1)_C \times SO(p_1)\times SO(p_2)\times \prod_\sigma U(r_\sigma)$, where $U(1)_{G}$, $U(1)_C$ arise from the dimensionally reduced metric $G_{9\mu}$ and antisymmetric tensor $C_{9\mu}$. 
As announced in the introduction, Eq.~(\ref{wls}) is telling us that Higgs and super Higgs mechanisms cancel each other for $F=1$, $a_\alpha=\half$, $a_\beta=0$ \ie for fermionic states stretched between the stacks of $p_1$ and $p_2$ $\half$-branes, which are in the bifundamental representation of $SO(p_1)\times SO(p_2)$. Notice that $a_\alpha={1\over 4}=-a_\beta$ also yields massless fermions. Hence, we expect Eq.~(\ref{vres}) to be valid when all WLs belong to $\{0,\half, \pm {1\over 4}\}$, even if there is a priori no reason to believe that $a=\pm {1\over 4}$ corresponds in general to critical points of~$\V$.  

The effective potential can be evaluated at 1-loop. When $M\ll M_s$, it takes the form~\cite{stableWL}
\be
\V={\Gamma(5)\over \pi^{14}}\, M^9\sum_{n_9}\frac{\N_{2n_9+1}(\W)}{(2n_9+1)^{10}}+{\cal O}\big((M_sM)^{9\over 2}e^{-\pi {M_s\over M}}\big),
\ee
where $\N_{2n_9+1}$ arises from  contributions of  the torus, Klein bottle, annulus and M\"obius strip amplitudes,  
\be
\begin{aligned}
\N_{2n_9+1}(\W)  &~=~4\big(\!-16-0-(\text{tr}\,\W^{2n_{9}+1})^{2}+\text{tr}\,(\W^{2(2n_{9}+1)})\big) \\
&~=~ -16\Bigg(\dis \sum_{\substack{r,s=1\\ r\neq s}}^{N}\cos\!\big(2\pi(2n_9+1)a_{r}\big)\cos\!\big(2\pi(2n_9+1)a_{s}\big)+N-4\Bigg),
\end{aligned}
\label{N9}
\ee
and where $N=16$ or 15 is the number of independent, dynamical WLs $a_r$. When all WLs are in $\{0,\half, \pm {1\over 4}\}$, we find as expected $\N_{2n_9+1} =n_F-n_B$, which yields Eq.~(\ref{vres}). Moreover, the WLs in the neighborhood of $a=\pm {1\over 4}$ have vanishing tadpoles only when $p_1=p_2$ \ie when the  configuration of $\half$-branes is symmetric with respect to the transformation $a\to \half -a$. However, restricting to the tachyon free configurations, WLs at $\pm {1\over 4}$ are excluded and only two solutions are found, namely $(p_1,p_2)=(32,0)$ and $(31,1)$ (up to the exchange $p_1\leftrightarrow p_2$). It turns out that both $SO(32)$ and  $SO(31)\times SO(1)$ configurations (where the inert $SO(1)$ reminds the presence of an isolated frozen $\half$-brane and fermions in the fundamental representation of $SO(p_1)$) yield a negative minimum of $\V$. However, we stress that the $SO(31)\times SO(1)$ solution has an energy slightly raised, as compared to the $SO(32)$ one. 


\section{Wilson line stability and positive potential in $\boldsymbol{d}$ dimensions}

By generalizing the above considerations to lower dimensions, the hope is that we may freeze more $\half$-branes at orientifold planes and obtain a non-negative potential. The type IIB orientifold theory compactified on $T^{10-d}$ can be analyzed in the ``most geometrical picture'' obtained by T-dualizing all of the internal directions. The internal space becomes a ``$(10-d)$-dimensional box'' of metric $\tilde G_{IJ}=G^{IJ}$, with one orientifold O$(d-1)$-plane at each of its $2^{10-d}$ corners. Moreover, the initial D9-branes turn into 32 $\half$-D$(d-1)$-branes, whose positions are $\tilde X^I=2\pi a_\alpha ^I\sqrt{\tilde G_{II}}$, $I=d,\dots, 9$, $\alpha=1,\dots, 32$. 

From the results  in 9 dimensions, we expect configurations where all $\half$-branes are coincident with O-planes to yield stable critical points of the potential, with respect to all WLs. Let us denote $p_A$ the number of $\half$-branes sitting on the $A$-th O-plane. By convention, we label the corners so that the coordinates of the  $(2A-1)$-th and $2A$-th differ only along the supersymmetry breaking direction $\tilde X^9$. In such a background, the numbers of massless bosonic and   fermionic degrees of freedom  turn out to be
\be
n_B =8\bigg(8+\sum_{A=1}^{2^{10-d}} \frac{p_{A}(p_{A}-1)}{2}\bigg)~,\;\;\quad n_F =8\sum_{A=1}^{2^{10-d}/2}p_{2A-1}p_{2A}.
\label{nfb}
\ee
$n_B$ contains $8\times 8$ states arising from the closed string sector, which correspond to the 10-dimensional dilaton $\phi$ and metric $G_{MN}$ in the NSNS sector, and the RR 2-form $C_{MN}$, $M,N=0,\dots 9$, which are dimensionally reduced. It also contain the bosonic parts of vector multiplets in the adjoint representations of the $SO(p_A)$'s gauge group factors, $A=1,\dots, 2^{10-d}$. On the other hand, $n_F$ contains the fermionic parts of vector multiplets in the bifundamental representation of $SO(p_{2A-1})\times SO(p_{2A})$, $A=1,\dots, 2^{10-d}/2$. They arise from the cancellation of the \SS and WL shifts of the momenta along the direction $X^9$ (in the original type I picture).

Notice that if Eq.~(\ref{vres}) holds when all  $\half$-branes are sitting on orientifold planes, this expression  is only the first term of a Taylor expansion in WLs. However, the linear term, when $\half$-branes move slightly, vanishes because the WLs are dressed by the charges of the states running in the virtual loop, and states with opposite charges can always be paired. At the next order, the mass terms of the WLs are proportional to squared charges. In general, WLs associated to a gauge group factor $\G_\kappa$ are non-tachyonic if~\cite{SNSM, CP}
\be
T_{{\cal R}^{(\kappa)}_B}-T_{{\cal R}^{(\kappa)}_F}\ge 0,
\ee
where $T_{{\cal R}^{(\kappa)}_B}$,  $T_{{\cal R}^{(\kappa)}_F}$ are the Dynkin indexes of the representations ${\cal R}^{(\kappa)}_B$ and ${\cal R}^{(\kappa)}_F$ of $\G_\kappa$ that are realized by  the $n_B$ and $n_F$ states. In our case, the above condition yields for $A=1,\dots, 2^{10-d}/2$
\be
\begin{cases}
p_{2A-1}-2-p_{2A}\ge~ 0, \quad \mbox{for the $SO(p_{2A-1})$ WLs,\;\;\; if $p_{2A-1}\ge 2,$}\\ 
p_{2A}-2-p_{2A-1}\ge 0, \quad\,\, \mbox{for the $SO(p_{2A})$ WLs,\,\;\;\;\;\quad  if $p_{2A}\ge 2$},
\end{cases}
\label{mas}
\ee 
whose compatibility  implies, up to exchanges $p_{2A-1}\leftrightarrow p_{2A}$, 
\be
\label{stabi}
\forall A=1,\dots, 2^{10-d}/2,\;\;\; (p_{2A-1},p_{2A})\mbox{ to be of the form} \,\begin{cases} 
\phantom{\mbox{or }}(p,0), \;\; p\ge 0\\ 
\mbox{or }(p,1),\;\; p\ge 3 \mbox{ or } p=1.
\end{cases} 
\ee

Without surprise, the $SO(32)$ configuration corresponding to all $\half$-branes coincident on a single orientifold plane yields the lowest value $n_F-n_B=-8\times 504$, and it is stable with respect to all open string WLs in arbitrary dimension $d$. The number of stable configurations and the rank of the gauge group fall, as we freeze $\half$-branes on O-planes in order to raise $n_F-n_B$. $\mbox{$\half$-branes}$ distributions yielding $n_F-n_B=0$ exist for $d\le 5$~\cite{stableWL}, with rank of the open string gauge group at most equal to 4. Among them, the simplest solutions involving the largest allowed values of $p_A$'s are
\be
\big[SO(5)\times SO(1)\big]\times\big[SO(1)\times SO(1)\big]^{13},\quad SO(4)\times\big[SO(1)\times SO(1)\big]^{14},
\ee
where the brackets indicate that the $\half$-branes are located on corners separated along $\tilde X^9$ only. The maximal value $n_F-n_B=8\times 8$ is reached when there is no left gauge symmetry arising from the open string sector, $[SO(1)\times SO(1)]^{16}$ for $d\le 5$. However, in all cases the models admit a $U(1)^{10-d}_G \times U(1)^{10-d}_C$ gauge symmetry arsing from the closed string sector, $G_{I\mu}$, $C_{I\mu}$, $I=d,\dots, 9$. 
The above results can also be derived by computing explicitly the 1-loop effective potential. Moreover, having  discussed so far non-tachyonic distributions of $\half$-branes, the potential will allow to conclude whether massless WLs introduce instabilities from interactions (still at 1-loop), an issue which arises for  the $SO(2)$ and $[SO(3)\times SO(1)]$ gauge group factors. 

In order to write the 1-loop potential~\cite{stableWL}, it is convenient to split the WLs as background values plus deviations, 
\be
a_\alpha^I = \langle a_\alpha^I \rangle +\varepsilon_\alpha^I, \quad \where \quad \langle a_\alpha^I \rangle\in\Big\{0,\half\Big\},\quad I=d,\dots, 9, \;\; \alpha=1,\dots, 32.
\ee
Note that the $\varepsilon_\alpha^I$'s can be arbitrary, and  their magnitudes are not supposed to be small. However, assuming that all mass scales in the undeformed background are greater than  $M$, the NSNS metric is bounded in the following sense, 
\be
G^{99}\ll  | G_{ij}|  \ll G_{99}, \quad |G_{9j}| \ll  \sqrt{G_{99}},\quad i,j=d,\dots,8 ,\quad G_{99}\gg 1.
\label{hyp}
\ee
The remaining moduli arise from the RR closed string sector and are the internal components $C_{IJ}$, $I,J=d,\dots, 9$, of the antisymmetric tensor. In these conditions, the potential reads
\be
\V=   {\Gamma\big({d+1\over 2}\big)\over \pi^{3d+1\over 2}}\, M^{d}\sum_{l_9}{\hat \N_{2l_9+1}(\varepsilon,G)\over |2l_9+1|^{d+1}} +\O\big((M_sM)^{d\over 2}e^{-2\pi c {M_s\over M}}\big),
\label{vtot}
\ee
where $c=\O(1)$ is positive. The dominant contribution is expressed in terms of 
\begin{align}
\label{hatn}
\hat \N_{2l_9+1}(\varepsilon,G) = &\; 4\,\bigg\{\!-16-0-\sum_{(\alpha,\beta)\in L}(-1)^{F}\cos\!\Big[2\pi(2l_9+1)\big(\varepsilon_\alpha^9-\varepsilon_\beta^9+{G^{9i}\over G^{99}}(\varepsilon_\alpha^i-\varepsilon_\beta^i)\big)\Big]\nonumber \\
 &  \qquad\qquad\qquad\qquad  \times ~\Hc_{d+1\over 2}\bigg(\pi|2l_9+1|{\big[(\varepsilon_\alpha^i-\varepsilon_\beta^i)\hat G^{ij}(\varepsilon_\alpha^j-\varepsilon_\beta^j)\big]^\half\over \sqrt{G^{99}}}\bigg)\\
& + \sum_\alpha  \cos\!\Big[4\pi(2l_9+1)\big(\varepsilon_\alpha^9+{G^{9i}\over G^{99}}\, \varepsilon_\alpha^i\big)\Big]  \Hc_{d+1\over 2}\bigg(2\pi|2l_9+1|{\big[\varepsilon_\alpha^i\,\hat G^{ij}\,\varepsilon_\alpha^j\big]^\half\over \sqrt{G^{99}}}\bigg)\bigg\},\nonumber 
\end{align}
which arises from the torus, Klein bottle, annulus and M\"obius strip amplitudes. In our notations, $L$ is the set of pairs $(\alpha,\beta)$ such that $\alpha$ and $\beta$ label $\half$-branes in the neighborhood of corners $2A-1$ and $2A$, for some $A=1,\dots,2^{10-d}/2$. Hence, for any $(\alpha,\beta)\in L$, massless strings stretched between these $\half$-branes contribute, and they generate the bosonic  adjoint and fermionic bifundamental representations of $SO(p_{2A-1})\times SO(p_{2A})$. Moreover,  we have defined 
\be
\begin{aligned}
&\hat G^{ij}= G^{ij}-{G^{i9}\over G^{99}}\, G_{99}\, {G^{9j}\over G^{99}} , \quad i,j=d,\dots, 8,\\
&\Hc_\nu(z)= {1\over \Gamma(\nu)}\int_0^{+\infty}{dx\over x^{1+\nu}}\, e^{-{1\over x}-z^2x}={2\over \Gamma(\nu)}\, z^\nu K_\nu(2z).
\end{aligned}
\ee
Some remarks are in order:

$\bullet$ Denoting the true dynamical degrees of freedom of the open string WLs as $\varepsilon_r^I$, $I=d,\dots, 9$, $r=1,\dots ,  \sum_{A=1}^{2^{10-d}} \lfloor {p_A/ 2}\rfloor$, $\hat \N_{2l_9+1}(\varepsilon,G)$ can be expanded to quadratic order to find, as expected,
\be
\begin{aligned}
\V&=\big(n_F-n_B\big)  \xi_d  M^d + \half\,  \xi_d^{\prime\prime}M^d\sum_r\left({p_{A(r)}-p_{\tilde A(r)}\over 2}-1\right)\varepsilon_r^I\hat \Delta_{IJ}\varepsilon_s^J\\
&\hspace{7cm}+\O(\varepsilon^4) +\O\big((M_sM)^{d\over 2}e^{-2\pi c {M_s\over M}}\big).
\end{aligned}
\label{vfinal}
\ee
In this expression, $A(r)$ denotes the corner around which the brane $r$ vary, while  $\tilde A(r)$ is the partner corner along the \SS direction $\tilde X^9$, while 
\be
\begin{aligned}
&\xi_d={\Gamma({d+1\over 2})\over \pi^{3d+1\over 2}}\sum_{n_9}\frac{1}{|2n_9+1|^{d+1}}, \quad \xi_d^{\prime \prime}={\Gamma({d+1\over 2})\over \pi^{3d+1\over 2}}\sum_{n_9}\frac{128\pi^2}{|2n_9+1|^{d-1}},\\
&\hat \Delta_{IJ}={1\over d-1}\!\left({G^{IJ}\over G^{99}}+(d-2){G^{I9}\over G^{99}}{G^{9J}\over G^{99}}\right).
\end{aligned}
\ee
It is not difficult to show that all eigenvalues of $\hat \Delta_{IJ}$ are strictly positive, so that the conditions~(\ref{mas}) for tachyons not to arise are recovered. 

$\bullet$ In particular, the WLs associated to $SO(2)$ and $[SO(3)\times SO(1)]$ gauge group factors are massless. However, it turns out the dominant term of the full 1-loop potential Eqs~(\ref{vtot}),~(\ref{hatn}) is totally independent of these WLs, which are therefore flat directions (up to the exponentially suppressed terms). 

$\bullet$ When Eq.~(\ref{stabi}) holds, keeping $M$ fixed, $\V$ is at a local minimum when all massive open string WLs are set to 0, while all massless ones are arbitrary. Thus,  Eq.~(\ref{vres}) remains valid in this more general case, since $[SO(3)\times SO(1)]$ can be broken by its WLs, with Higgs masses lower than $M$. Furthermore, the minima are independent of the NSNS moduli $\hat G^{ij}$, $G^{i9}$, $i,j=d,\dots 8$, and of  the RR ones $C_{IJ}$, $I,J=d,\dots, 9$.  Therefore, all moduli arising from the closed string sector are flat directions, except $M$ (unless $n_F-n_B=0$).


\section{Conclusion and remarks}

We have seen that at fixed supersymmetry breaking scale $M$, and up to exponentially suppressed corrections, local minima of the 1-loop effective potential of arbitrary sign exist in dimension $d\le 5$, in type I string theory. In a T-dual picture, they are realized by freezing  isolated $\half$-branes on orientifold planes, whose effect is to reduce the rank of the gauge symmetry and to introduce massless fermions in the fundamental representations of $SO(p)$ gauge group factors. Therefore, the WL matrices 
\be
\W_I=\diag(e^{2i\pi a^I_\alpha}, \alpha=1,\dots, 32), \quad I=d,\dots,9,
\ee
can be elements of $SO(32)$,  or in $O(32)$ but not in $SO(32)$. Many of the configurations we have described correspond to the former case, and heterotic dual descriptions should exist. However, when WL matrices are not in $SO(32)$, no heterotic dual description can be defined and inconsistencies at the non-perturbative level are expected to arise~\cite{Witten:1998cd}.


\section*{Acknowledgement}
 
This work  was partially supported by the Royal Society International Cost Share Award. 


\end{document}